# Universal radiation dynamics by temporal transitions in optical waveguides

Amir Shlivinski[1] and Yakir Hadad[2*]

[1]*School of Electrical and Computer Engineering, Ben-Gurion University, Beer Sheva, Israel*

[2]*School of Electrical Engineering, Tel-Aviv University, Tel-Aviv, Israel 69978.*

*\*hadady@eng.tau.ac.il*

**Abstract:** When an excited electromagnetically open optical waveguide goes through a temporal transition of its material properties, it radiates to the ambient surroundings. In this letter, we explore this radiation and reveal, using asymptotic evaluation of path integral in the complex frequency (Laplace) plane, a peculiar space-time dependence of its frequency. Specifically, we derive an exact formula (Eq. (11)) for the instantaneous radiation frequency, which exhibits a chirp behavior with respect to time. This simple formula depends on the ambient properties and on the longitudinal wavenumber $\beta$ of the guided mode before the temporal transition but not on the specific waveguide structure or materials. In addition, we derive a $t^{-3/2}$ decay rate of the radiative field on time. We verify our analytic results using full-wave simulations of a dispersive and lossy Indium Tin Oxide waveguide that undergoes smooth temporal long transitions over up to ~200 cycles at the initially guided mode frequency. Thus, these theoretical findings offer valuable insights into the behavior of general optical waveguides experiencing temporal transitions and provide a powerful tool for analyzing and designing such THz and optical setups, with potential use in sensing and imaging.

*Introduction.* - Temporal interfaces have been extensively investigated since late 1950s [1-4], in the context of magnetoelastic waves [5-6], in transmission line systems [7-8], for signal manipulations, such as stretching, compression, and time-inversion [6-8], and in plasma [9-10]. In recent years, the field has experienced a resurgence of interest [11-16] due to extensive research on wave manipulations by metamaterials and advancements in technology, enabling practical high-rate temporal modulations (e.g., see [17,18]). The integration of temporal variation of the guiding medium, often referred to as the "fourth dimension" for wave design [19], has garnered significant attention. This novel approach has shown promise in surpassing known performance bounds and revealing intriguing wave phenomena, and functionalities [20-36].

While temporal interfaces have been extensively studied theoretically, experimental demonstrations are less common. Implementing abrupt changes in the medium parameters on a large scale poses significant challenges. However, practical implementations of temporal discontinuities appear more feasible when confined to compact domains such as cavities [35-39], metasurfaces [40-45], thin dielectric layers [25,46], and transmission lines [7, 8, 12, 21, 34, 47]. For instance, the authors in [8,46] demonstrated experimentally signal compression and inversion by temporal discontinuities in transmission lines at radio frequencies, demonstrating effects such as signal stretching and compression, and time-reversal. In the THz and optical frequencies, transmission lines are impractical, and instead, thin dielectric layers and interfaces acting as waveguides can be used [18, 46]. For instance, Bohm et al. [18] demonstrated optical switching of epsilon-near-zero plasmon resonances in Indium Tin Oxide (ITO) layer, and, Miyamaru et al. [46] experimentally demonstrated ultrafast frequency shifts in the THz range caused by a time-varying dielectric layer. Thus, providing the first steps towards practical potential implementations for temporal discontinuities in the optical regime.

In this letter, we focus on the radiation dynamics by a temporal transition of an optical waveguide. To that end, we apply asymptotic evaluation of complex path integrals about the branch cut singularity of a spectral integral that represents the radiation dynamics. This procedure yields a closed-form formula for the *radiation wave* caused by the temporal transition. Intriguingly, we show that the radiation frequency possesses a chirp-like dependence on time. We show that this dependence is universal, independent of the specific waveguide geometry and materials, but only depends on the initial mode longitudinal wavenumber and the ambient wave velocity.

*Case study: Temporally switched dielectric waveguide.* - Inspired by the experimental setups in [18,46], we choose to examine the configuration depicted in Fig. 1(a). We assume that the waveguide consists of a general dispersive dielectric that follows a typical Drude model with $\epsilon_{d_1} = \epsilon_{\infty 1} - \omega_{p1}^2/(\omega^2 - j\omega\Gamma_1)$ where $\epsilon_{\infty 1}$, $\omega_{p1}$ and $\Gamma_1$ denote the model parameters, high-frequency permittivity, plasma frequency, and damping rate, respectively. The waveguide thickness is $d$ and is backed by a perfect magnetic conductor (PMC), implying that in a realistic setup, the system is assumed symmetric. This specific choice does not limit the validity of our result regarding the temporal chirped dynamics of the radiation frequency, which apply to any planar waveguide. The waveguide supports a guided mode, propagating in the $+z$ direction, with wavenumber $\beta$. Thus, the fields can be expressed by

$$\boldsymbol{E}_1^{a,d}(\boldsymbol{r},t) = \boldsymbol{e}_1^{a,d}(x,t)e^{-j\beta z}, \qquad \boldsymbol{H}_1^{a,d}(\boldsymbol{r},t) = \boldsymbol{h}_1^{a,d}(x,t)e^{-j\beta z} \qquad (1)$$

where superscript $a, d$ stand for 'air' and 'dielectric'. This initial field is time-harmonic with time-dependence $e^{j\omega_0 t}$. At $t = 0$ an abrupt temporal transition of the dielectric material properties takes place (smooth transitions are considered later). At $t > 0$ we assume to have new parameters in the Drude model: $\epsilon_{\infty 2}, \omega_{p2}$, and $\Gamma_2$. Consequently, the fields are altered. We denote the fields at $t > 0$ by $\boldsymbol{E}_2^{a,d}(\boldsymbol{r},t)$ and $\boldsymbol{H}_2^{a,d}(\boldsymbol{r},t)$. Note that the new fields are not time-harmonic anymore. Instead, they consist of various wave components, guided modes, and radiation waves with different time dependencies and frequencies (see illustration in Fig. 1(b)). In order to find the fields after the temporal discontinuity, we apply the unilateral Laplace transform, as previously used e.g., in [44,48]. The detailed derivation for our problem is provided in the Supplementary material file [49]. Briefly, since the waveguide dielectric is assumed dispersive, we have to impose the following continuity across the temporal boundary at $t = 0$ to obtain the fields at $t = 0^+$,

$$Y_1^{a,d}(r,t^{0-}) = Y_2^{a,d}(r,t^{0+}), \tag{2}$$

where, for non-magnetic dielectric material $Y$ stands for the electric field $E$, magnetic induction $B$, polarization density $P$ and its time-derivative $\partial_t P$ in the air domain and in the dielectric slab. Eq. (2) enforces that the $z$ behavior of the fields is preserved (conservation of quasi-momentum in $z$) through the temporal boundary, thus implying that the $z$ dependence of $E_2^{a,d}$ and $H_2^{a,d}$ is also $e^{-j\beta z}$ and as a result $\partial_z \to -j\beta$ in Maxwell's equations [49] (Sec. S1B Eq. (S6)). The fields at $t = 0^+$, which are obtained by the continuity conditions in Eq. (2), serve as spatial initial conditions for a solution of the Helmholtz equation formulated in the Laplace domain for the wave-fields amplitude at t > 0. The latter reads,

$$\partial_x^2 \tilde{h}_z(x,s) - \gamma^2 \tilde{h}_z(x,s) = F(x,t=0^+;s) \tag{3}$$

where $\tilde{h}_z = \tilde{h}_z^{(a,d)}$, and $\gamma = \gamma_{(a,d)}$. The right-hand side of Eq. (3) is given explicitly in the supplementary material file [49] (see Eqs. (S10) and Eq. (S37) there), together with $\gamma_d$, which possesses a relatively complex mathematical structure (Eq. (S15) and Eq. (S46) in [49]). Since we are interested in the radiation into the air domain, the main focus of the discussion is $\gamma = \gamma_a$ that reads,

$$\gamma = \sqrt{s^2/c^2 + \beta^2} \tag{4}$$

and $c$ is the speed of light in a vacuum (used for the air domain). The solution of $\tilde{h}_z(x,s)$ in Eq. (3) is provided in Eq. (S47-S48) in [49]. Next, the inverse Laplace transform is applied to obtain the time domain fields,

$$h_z^{a,d}(x,t) = \frac{1}{2\pi j} \int_{-j\infty+\sigma}^{j\infty+\sigma} \tilde{h}_z^{a,d}(x;s) e^{st} \, ds. \tag{5}$$

The physical realization of the switching process dictates that following the abrupt switching of the slab's dielectric constant, a set of orthogonal surface wave modes of the medium are excited in additional to transient radiation (continuous spectrum). This is readily observed in the Laplace formulation. Specifically, the spectral function $\tilde{h}_z^{a,d}(x;s)$ contains two types of singularities. Isolated poles, and branch points and corresponding branch-cuts (for further details, refer to Sec. S1d in [49]). See Fig. 1(c) for illustration. Formally, the integration path parameter $\sigma > 0$ should be selected such that the integration is performed in the transform region of convergence, namely, since our problem is causal and stable, to the right of the imaginary axis on the $s$-plane. However, to calculate the fields at $t > 0$ this formal integration path can be deformed to encircle the singular points in the complex $s$ plane. Thus, we get deeper physical insight on the various wave species that are excited during the temporal discontinuity. Specifically, the pole singularities, coming in conjugate pairs, correspond to the counter propagating guided modes in the $\pm z$ directions after the temporal switching. Moreover, besides the poles, there are additional two branch point singularities that originate from the square roots in the complex propagation constant along the $x$ direction, in the air and in the dielectric, $\gamma_a$ and $\gamma_d$, respectively. However, it can be readily verified that due to the mirror symmetry of the solution in $x$, $\tilde{h}_z(x,s)$ is even in $\gamma_d$, and therefore $\gamma_d = 0$ is not a branch point of the spectral field. This leaves only the branch points due to $\gamma_a$. The integration around the branch point provides a continuous frequency spectrum of waves that are excited due to the temporal discontinuity and radiated into the air domain. Their behavior is elaborated below and is the key result of this letter.

*Derivation of the radiative field.* - Our analysis begins by writing the integral around the branch cut with branch point at $s_{bp}$ (upper branch in Fig. 1(c)). The integration takes the following canonical form

$$I_{bc}^U = \int_{bc} f(s)e^{\psi(s)}ds \tag{6}$$

where $f(s)$ is a slow varying function in $s$. For example for the $E_y$ field component it reads,

$$f(s) = \frac{1}{2\pi j}\frac{\mu_0 s D(s)}{\gamma_a(s)}$$

and it depends on the field's initial conditions as well as on the boundary and the time continuity conditions (see the Eqs. (S17) and (S47) in [49] for the expression of $D(s)$). Furthermore, $\psi(s)$ in Eq. (6) is given by

$$\psi(s) = -\gamma_a(x-d) + st \tag{7}$$

denoting the phase term. As opposed to $f(s)$ which is slowly varying and independent of the physical coordinates $x$ and $t$, the exponential term in Eq. (6) is highly oscillatory except in the vicinity of a stationary point, if exists. The integration in Eq. (6) is carried around the upper branch cut as shown in Fig. 1(c). In this case, the variable $s$ passes in the range $(i\infty + \epsilon, s_{bp} + \epsilon) \cup (s_{bp} - \epsilon, i\infty - \epsilon)$ where $\epsilon \to 0^+$. Along this $s$-path, the variable $\gamma_a$ maps into the more convenient range $(i\infty, -i\infty)$. Therefore, we perform the following change of integration variable: $s \to \gamma_a$. Recall Eq. (4), from which we have $ds = d\gamma_a\, c^2\gamma_a/s$. By plugging the above change of variables into Eq. (6) we get,

$$I_{bc}^U = \int_{i\infty}^{-i\infty} \frac{c^2\gamma_a}{s(\gamma_a)} f(s(\gamma_a))e^{\psi(s(\gamma_a))}d\gamma_a \tag{8}$$

For a remote observer, $x \gg \lambda$ and long enough time after the causal time at which the wave front reaches the observer at $x$, the integrand in Eq. (8) possesses a stationary phase point behavior as shown in Fig. 2(ai)-(aiii). The stationary phase point is found by solving $\psi'(\gamma_a) = 0$. Which leads to

$$\gamma_{a,s} = \frac{\beta(x-d)}{\sqrt{(x-d)^2 - (ct)^2}}. \tag{9}$$

This stationary point is imaginary and thus located along the integration path only in causal times, i.e., for $ct > x - d$. Asymptotically, the contribution to the branch cut integral is, therefore, dominated by the stationary point contribution and takes the following form [50, Ch. 4],

$$I_{bc} \sim \sqrt{-j2\pi \frac{\omega_s^3(x,t)}{\omega_{bp}^2 t} \frac{c^2 \gamma_{a,s}}{s(\gamma_{a,s})}} f(s(\gamma_{a,s})) e^{\psi(s_s)} H\left(t - \frac{x-d}{c}\right). \tag{10}$$

Where $H(\cdot)$ denotes the Heaviside step function, representing the fact that the wavefront reaches the observer at the casual time $t_{causal} = (x-d)/c$, and $\omega_s(x,t)$ as well as $\omega_{bp}$ are given in Eq. (11) below. Using Eq. (9), it is straightforward to show by Eq. (10) that the radiative field exhibits an asymptotic time dependence $\sim t^{-3/2}$ for a fixed observer location $x$.

*Chirped-like radiation frequency.* - In order to identify the radiation frequency due to the temporal transition we resort back into the $s$ variable that is directly connected to the temporal frequency. Using Eq. (4) with Eq. (9) we have

$$\omega_s(x,t) = js_s = \frac{\beta c^2 t}{\sqrt{(ct)^2 - (x-d)^2}} \tag{11}$$

Eq. (11) is the main result of this letter. It implies that the radiation frequency depends on space and time and exhibits a chirp-like *behavior*. For a fixed observer location $x$, and long time after the wavefront arrival time ($t_{causal}$), the radiation frequency settles at $\omega_s \to \beta c = js_{bp} = \omega_{bp}$. However, shortly after the wavefront arrival time, the frequency exhibits a singular behavior as the denominator in Eq. (11) vanishes. Mathematically, this result results from the stationary phase point motion as time progresses. This behavior is shown in Fig. 2(ai)-(aiii) for three time-instances, $t = (3,10,50)t_{causal}$, respectively. All the numerical results in this letter are shown for a layer of thickness $d = 0.2\lambda_0$, where $\lambda_0 = 2\pi c/\omega_0$ where $\omega_0 = 2\pi \cdot 120[\text{THz}]$ is the excitation frequency of the guided modes prior to the temporal transition, and with Drude model parameters: $\omega_{p1} = 1376\text{ rad} \cdot \text{THz}$, $\Gamma_1 = 147\text{ rad} \cdot \text{THz}$, $\epsilon_{\infty 1} = 4.31$ (the model values were taken from [51]) before the transition, and a 15% higher plasma frequency $\omega_{p2} = 1582\text{ rad} \cdot \text{THz}$ following the transition. The resulting time-frequency behavior is shown in a spectrogram given in Fig. 2(b). The spectrogram that is calculated for the exact solution is compared with the curve obtained by the asymptotic result in Eq. (11) (dashed black line). An excellent agreement between the results is evident. Remarkably, as will be shown below, using finite-difference-time-domain (FDTD) simulations, this chirp behavior also holds nicely for gradual, highly smooth temporal transitions.

*"Quantum mechanical" interpretation of the radiation frequency.* - The derivation of Eq. (11) above required extensive analytical work. Here, using simple kinematic arguments we provide an alternative derivation, that while not strictly formal, provides some physical insight into the peculiar time-frequency behavior of Eq. (11). To that end, consider a radiative photon emitted at the temporal switching time $t = 0$, from a point $-z_{ph}$ along the interface. We denote by $t$ the time it propagates before reaching the observer at $z = 0$ and $x_{ph} = x - d$. Thus, the photon passes along a distance equals to $ct$, which means that it passes along parallel to the interface distance that equals to $z_{ph} = \left((ct)^2 - x_{ph}^2\right)^{0.5}$. See Fig. 2(c) for illustration. On the other hand, along the $z$ direction we have conservation of momentum during the switching process. Using De Broglie hypothesis, the photon quasi momentum component along $z$ equals $p_{z,ph} = \beta\hbar$, while the photon effective mass reads $m_{eff} = E/c^2 = \hbar\omega/c^2$. Thus,

$$z_{ph} = v_{z,ph} t = \frac{p_{z,ph}}{m_{eff}} t. \qquad (12)$$

By plugging the expressions for the quasi-momentum and the effective mass, we immediately obtain Eq. (11) above. Obviously, this quantum-like derivation should be demystified by noting that the definition of a photon mass here should be considered only as an elegant manifestation to of the phase velocity, and its connection to that of the guided mode before the temporal switching. Yet, this simplistic derivation captures nicely a complicated wave physics.

*Gradual temporal transitions.* - In this section, we demonstrate by full-wave FDTD simulations (using in-house developed code, using [52]) that the theoretical prediction of the radiation frequency in Eq. (11) also holds for the case of gradual temporal transitions. In Fig. 3(a), we provide the transition profile. This profile is applied to the plasma frequency. Thus, we have in our FDTD simulations $\omega_p(t)$ – that is varying continuously in time. For simplification, we assume that the other parameters of the Drude model, $\Gamma$ and $\epsilon_\infty$, remain unchanged. The profile function in Fig. 3(a) represents the transition between the two material states, 1 and 2. In the general case, the transition process involves a rise time $T_r$ from state 1 with $\omega_{p1}$ to state 2 with $\omega_{p2}$, followed by a slower relaxation process with a fall-time $T_f$ from state 2 back to 1. The Drude model parameters used in these simulations are identical to those used to calculate Fig. 2. In Fig. 3(b), only a single transition occurs from state 1 to state 2. The rise time is taken to be $T_r = 10T_0$ when $T_0$ denotes the period of the guided mode before the transition starts; thus, $T_0 = 2\pi/\omega_0 = 8.33$ [fs]. In this example, we consider no relaxation, and thus, $T_f = \infty$. The field spectrogram at $x = 90\lambda_0$ shows the chirped radiation frequency behavior with an excellent agreement to the formula in Eq. (11). As expected, the wavefront reaches the observer at the causal time. At that instant, the radiation frequency is very high and gradually decreases over time to the steady radiation frequency corresponding to the branch point. Figs. 3(c-d) show results with a more realistic transition involving a slower temporal transition, $T_r = 50T_0 = 0.42$ [ps], and relaxation time $T_f = 150T_0$, thus a total transition duration of about 200 cycles. The time axis of Fig. 3(a) corresponds quantitatively to this case. In Fig. 3(c) the electric field at $x = 90\lambda_0$ is observed, clearly demonstrating the theoretically anticipated $t^{-3/2}$ decay. In Fig. 3(d) a spectrogram of the field (as in Fig. 3(b)) is shown, and an excellent agreement is obtained by comparison to the theoretically anticipated frequency-time dependence in Eq. (11) that is shown by the black-dashed curve.

*Conclusions.* - This letter explored the radiation dynamics due to temporal transition in an electromagnetically open waveguide. Our approach provides a comprehensive understanding of the wave dynamics during the temporal switching. In particular, our approach sheds light on the unique radiation characteristics, exhibiting universal radiation frequency behavior independent of the specific waveguide configuration as given by Eq. (11) above, and with an asymptotic $t^{-3/2}$ decay rate. These results hold firmly also in the case of smooth temporal transitions as validated for realistic gradual transitions in Fig. 3. As such, our quantitative description of the radiation dynamics can be used as a unique fingerprint applied in far-field sensing and characterization of temporal transitions in nanoscale THz and optical devices, as well as a new venue to explore optical non-equilibrium processes as studied in [53].

**Acknowledgment**

This research was supported by the ISRAEL SCEINCE FOUNDATION (grant #1457/23).

**Figures**

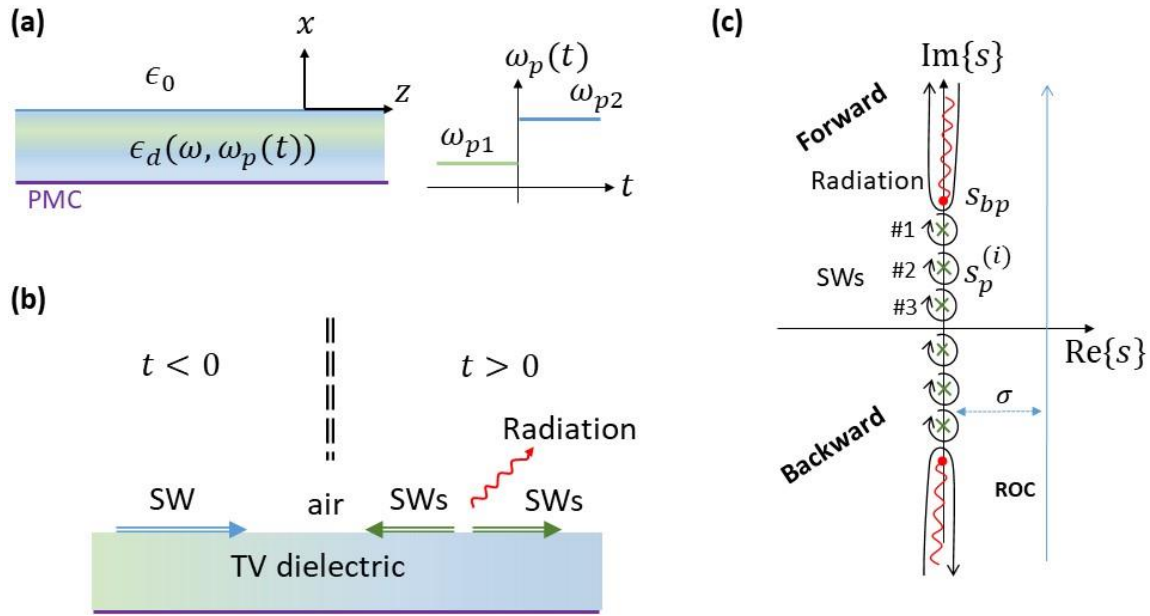

**Figure. 1.** (a) Time-varying dielectric slab optical waveguide. The waveguide is surrounded by air with $\epsilon \approx \epsilon_0$ while the slab layer consists of a dispersive time-varying dielectric material that its plasma frequency changes between two values, $\omega_{p1}$ and $\omega_{p2}$ at time $t = 0$. A perfect magnetic conductor backs the layer. (b) At $t < 0$ a TE Surface Wave (SW guided mode) is propagating. At $t > 0$, after the temporal discontinuity forward and backward SWs as well as radiation is expected. (c) The complex $s$ plane. Poles correspond to guided modes that may be excited after the switching, while the branch cuts correspond to radiation waves.

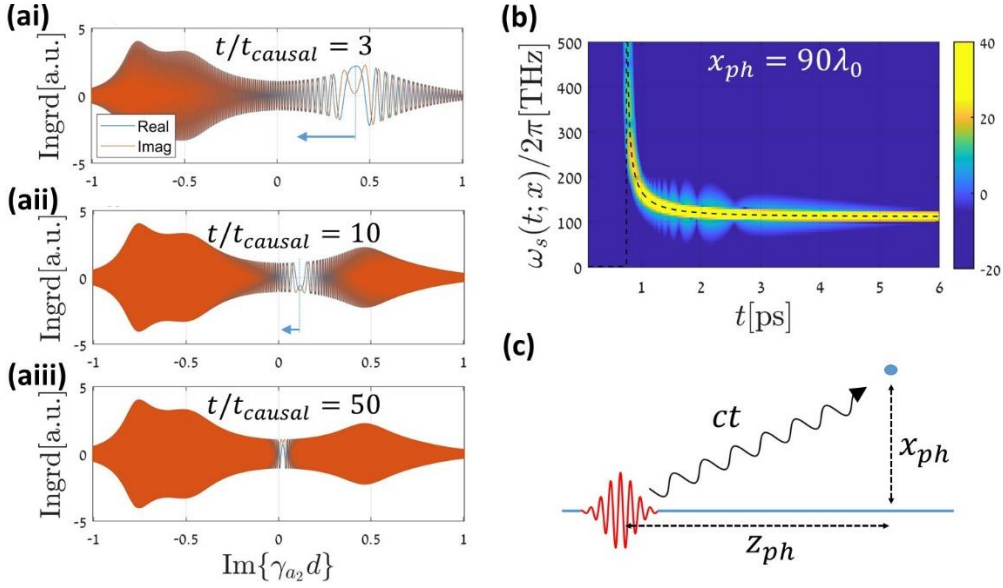

**Figure 2.** The observed radiation frequency. (ai) Numerical demonstration of the stationary phase for a fixed distance $x_{ph} = 90\lambda_0$ from the interface, and at a fixed time $t = 3t_{causal}$, not too long after the causal time $t_{causal} = x_{ph}/c$. (aii,aiii) As (ai) but for a longer time after the temporal switching, $t = 10t_{causal}$, and $t = 50t_{causal}$. The stationary point is more localized and gets closer to the branch point at $\gamma_a = 0$ as the time elapses. (b) A spectrogram picture, frequency vs. time, for an observer located at $x_{ph} = 90\lambda_0$. The dominant frequency varies with time and precisely follows the asymptotic formula for the radiation frequency in Eq. (10). (c) A "quantum mechanical" interpretation for the stationary phase result. The photon here is treated as a particle.

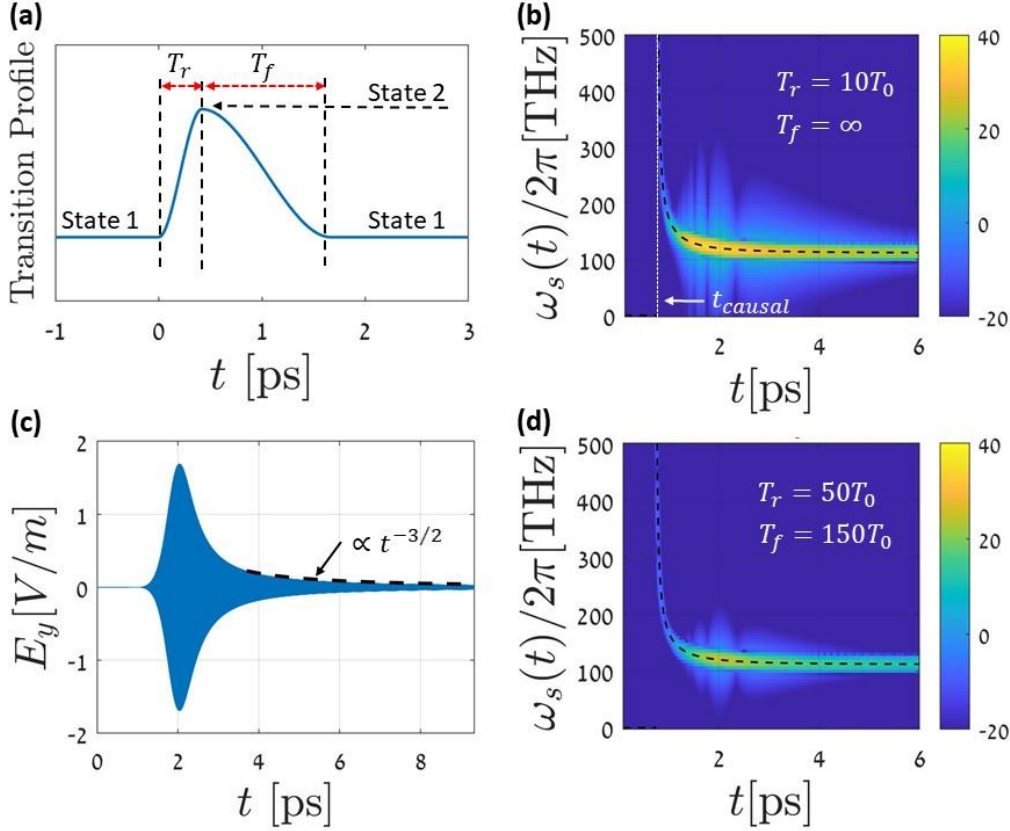

**Figure 3.** (a) The transition profile of the plasma frequency. (b) Spectrogram of the FDTD simulation field at $x = 90\lambda_0$ with transition rise time $T_r = 10T_0$ and infinite fall time (i.e., a transition from state 1 to state 2). The wavefront arrival time is marked by $t_{causal}$. The analytic result in Eq. (11) is shown by the black dashed line, indicating in excellent agreement with the gradual transition case. (c-d) Simulation results with an even slower transition $T_r = 50T_0$ and with relaxation $T_f = 150T_0$ (state 1 to 2 and back to 1 – see (a)). The observer is located at $x = 90\lambda_0$. (c) The electric field at $x = 90\lambda_0$. The field oscillations are too dense to be discerned in this scale. The envelope asymptotically decays as $t^{-3/2}$. (d) As (b) but for the transition scenario in (c).